\documentclass{appolb}
\usepackage{graphicx}
\usepackage{amsmath}
\usepackage{ae} 			
\usepackage[english]{babel} 	

\begin{document}
\title{Towards a consistent description of in-medium parton branching %
\thanks{Presented at Excited QCD 2015}%
}
\author{Liliana Apolin\'{a}rio, Guilherme Milhano
\address{CENTRA, Instituto Superior T\'{e}cnico, Universidade de Lisboa, Av. Rovisco Pais, P-1049-001 Lisboa, Portugal}
\newline
\newline
{N\'{e}stor Armesto, Carlos A. Salgado
}
\address{Departamento de F\'{i}sica de Partículas and IGFAE, Universidade de Santiago de Compostela, E-15706 Santiago de Compostela, Galicia-Spain}
}
\maketitle
\begin{abstract}
Ultra-relativistic heavy-ion collisions are a window of opportunity to study QCD matter under extreme conditions of temperature and density, such as the quark-gluon plasma. Among the several possibilities, the study of jet quenching - generic name given to in-medium energy loss modifications of the parton branching - is a powerful tool to assess the properties of this new state of matter. The description of the parton shower is very well understood in vacuum (controlled reference) and medium-induced modifications of this process can be experimentally accessed through jet measurements. Current experimental data, however, cannot be entirely described only with energy loss phenomena. TTransverse momentum broadening and decoherence effects, both theoretically established by now, and their interplay are essential to build a consistent picture of the medium-modifications of the parton branching and to achieve a correct description of the current experimental data. In this write-up, we will present the latest developments that address such unified description.
\end{abstract}
\PACS{PACS numbers come here}
  
\section{Introduction}

\par The assessment of the properties of the quark-gluon plasma (QGP) that is formed in ultra-relativistic heavy-ion collisions at the LHC is at the forefront of current efforts of the heavy-ion physics program. The wide range of scales available to probe this hot and dense medium allows the characterisation of this matter using different sectors of the QCD theory. As such, it is also a perfect laboratory to test the frontier of our knowledge of theory of the strong interactions. Among the large diversity of hard probes, jets - spray of collimated particles - are the ideal objects to probe the medium modifications of the QCD partonic branching since they are less dependent on hadronization models. In vacuum, the evolution of the parton shower from a large virtuality scale down to a lower cut-off scale is successfully described by perturbation theory. In the presence of a hot, coloured and dense medium, however, modifications generically known under the name of \textit{Jet Quenching} that include additional energy loss processes, are expected to occur. In the following sections, the main in-medium modifications in the description of the parton shower with respect to vacuum, within a perturbative approach, will be briefly overviewed.

\section{Path-integral approach to jet quenching}

Particles produced in the hard scattering are extremely energetic when compared to the usual energy scale of the medium constituents through which it travels. As such, making use of a high-energy approximation, it is assumed that the particle undergoes multiple soft scatterings with the medium that results into a momentum kick in the transverse direction plus a colour field rotation, without degrading its longitudinal energy\footnote{Light-cone coordinates will be used throughout the manuscript whose relation with Minkowski coordinates is given by: $x_\pm = (x_0 \pm x_3)/\sqrt{2}$ and $\mathbf{x} = (x_1, x_2)$. Moreover, the gauge is fixed such that the color field component $A_+ = 0$.}. Formally, such propagation of a particle with longitudinal momenta $p_+$ from $x_{0+}$ at transverse position $\mathbf{x_0}$ to $L_+$ at $\mathbf{x}$ is described by a Green's function:
\begin{equation}
\begin{split}
	G (x_{0+}, \mathbf{x_{0}}; L_+, \mathbf{x} | p_+) = & \int_{\mathbf{r}(x_{0+}) = \mathbf{x_{0}}}^{\mathbf{r} (L_+) = \mathbf{x}} \mathcal{D} \mathbf{r} (\xi) \exp \left\{ \frac{ip_+}{2} \int_{x_{0+}}^{L_+} d\xi \left( \frac{d \mathbf{r}}{d\xi} \right)^2 \right\} \\
	& \times W(x_{0+}, L_+; \mathbf{r}(\xi)) \, ,
\end{split}
\end{equation}
whose Wilson line is path-ordered:
\begin{equation}
	W (x_{0+}, L_+; \mathbf{r}(\xi)) = \mathcal{P} \exp \left\{ i g \int_{x_{0+}}^{L_+} d\xi A_- (\xi, \mathbf{r}(\xi)) \right\} \, .
\end{equation}

\begin{figure}[htb]
\centerline{%
\includegraphics[width=0.5\textwidth]{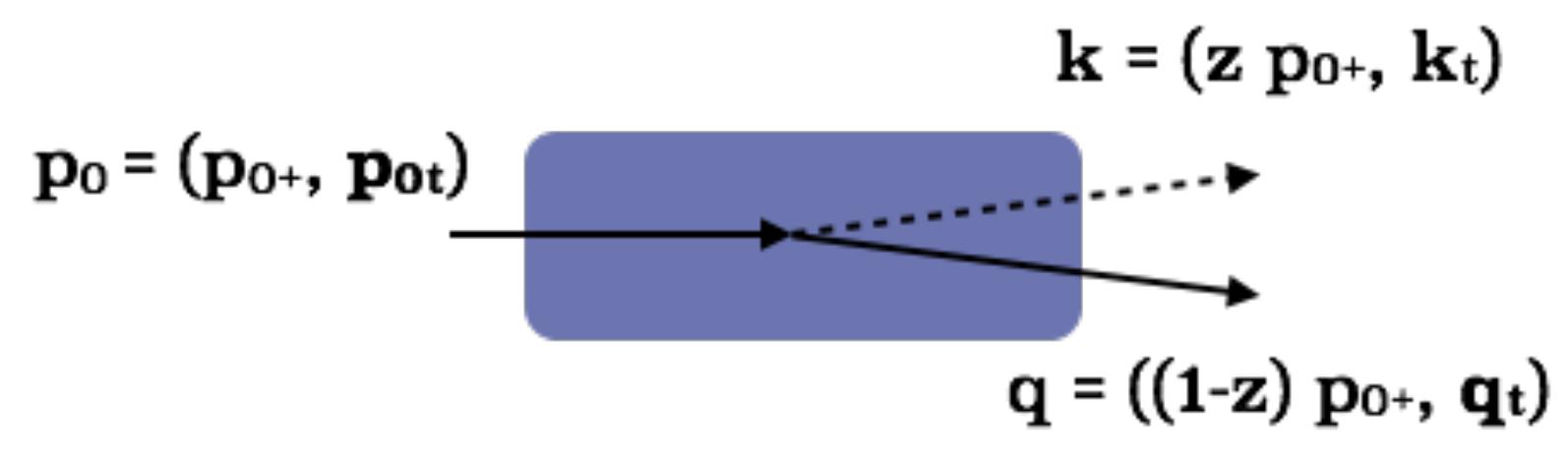}}
\caption{Schematic view of the bremsstrahlung process in the presence of a finite hot and dense medium, represented as a shaded area. The solid lines represent a quark while the dashed lines a gluon that is emitted with a fraction $x$ of the parent parton plus momentum $p_{0+}$.}
\label{Fig:amp}
\end{figure}

The calculation of the in-medium gluon bremsstrahlung process, within perturbation theory, accounts for the evaluation of the elementary gluon emission vertex inside a finite medium (see figure \ref{Fig:amp}). Within the employed approach, the medium is seen as a collection of static scattering centres whose colour field configuration is frozen during the propagation time of the fast particle. As such, to account for the total cross-section, an average over the possible ensemble of medium configurations has to be performed\footnote{Details of such calculations can be found in \cite{CasalderreySolana:2007zz,Apolinario:2014csa}}. For that, a factorisation between the acquired Brownian motion and the color field rotation is usually assumed.

\section{Jet quenching: latest developments}

Recent results from the several LHC collaborations have shown that \cite{Aad:2010bu,PhysRevC.84.024906}: (i) jets lose a significant fraction of its initial energy; (ii) its azimuthal direction is mildly modified with respect to vacuum ; (iii) the lost energy can only be recovered at very large distances away from the dijet axis in the form of soft particles. Current jet quenching Monte Carlos \cite{Armesto:2009fj,Zapp:2012ak} based on phenomenological extensions of soft in-medium gluon radiation\cite{Baier:1998kq,Zakharov:1997uu}, have shown that, within current experimental uncertainties, they are able to describe fairly well such observables. Nonetheless, when compared to experimental results on energy and particle distribution inside a jet (jet fragmentation functions)\cite{Aad:2014wha,PhysRevC.90.024908}, such models fail to describe the main observed features: the core of the jet is unmodified with respect to vacuum while the outer layers are largely affected by the medium. 

Such fact motivated several recent theoretical improvements of jet quenching models based on a path-integral approach, to go beyond the eikonal approximation that was usually employed. These include a better description of energy loss processes \cite{Apolinario:2012vy,Ovanesyan:2011kn,D'Eramo:2010ak}, momentum broadening \cite{Blaizot:2012fh} and decoherence effects \cite{MehtarTani:2010ma}. In particular, it was shown in \cite{Blaizot:2012fh} that, in the limit of an infinite medium, the in-medium double differential gluon bremsstrahlung process can be seen as a collection of "quasi-local" emissions, i.e., factorization of the independent broadening of each final particle that participates in the gluon emission vertex being the kinematic of the vertex described by the usual vacuum QCD splitting function.
%
Considering the QCD antenna-setup in the eikonal approximation ($z \rightarrow 0$), it was shown in \cite{MehtarTani:2010ma} that the medium opens the available phase space for radiation. In comparison with vacuum calculations, in addition to the usual angular ordering feature, an additional "anti-angular" component was found for the total number of gluons that is emitted from one of the emitters from the atenna:
\begin{equation}
	dN_q = \frac{\alpha_s C_F}{\pi} \frac{dk_+}{k_+} \frac{\sin \theta d\theta}{ 1 - \cos \theta} \left[ \Theta(\cos \theta - \cos \theta_{q\bar{q}}) + \Delta_{med} \Theta ( \cos \theta_{q\bar{q}} - \cos \theta) \right] \, ,
\end{equation}
weighted by a $\Delta_{med}$ parameter,
\begin{equation}
	\Delta_{med} \approx 1 - \exp \left\{ - \frac{1}{12} \hat{q} \theta_{q\bar{q}} L^3 \right\} \, ,
\end{equation}
that depends on the medium length, $L$, and on the angle between the quark-antiquark legs that form the antenna, $\theta_{q\bar{q}}$. For a very opaque medium, the "anti-angular" ordering contribution is maximum.

\begin{figure}[htb]
\centerline{%
\includegraphics[width=0.6\textwidth]{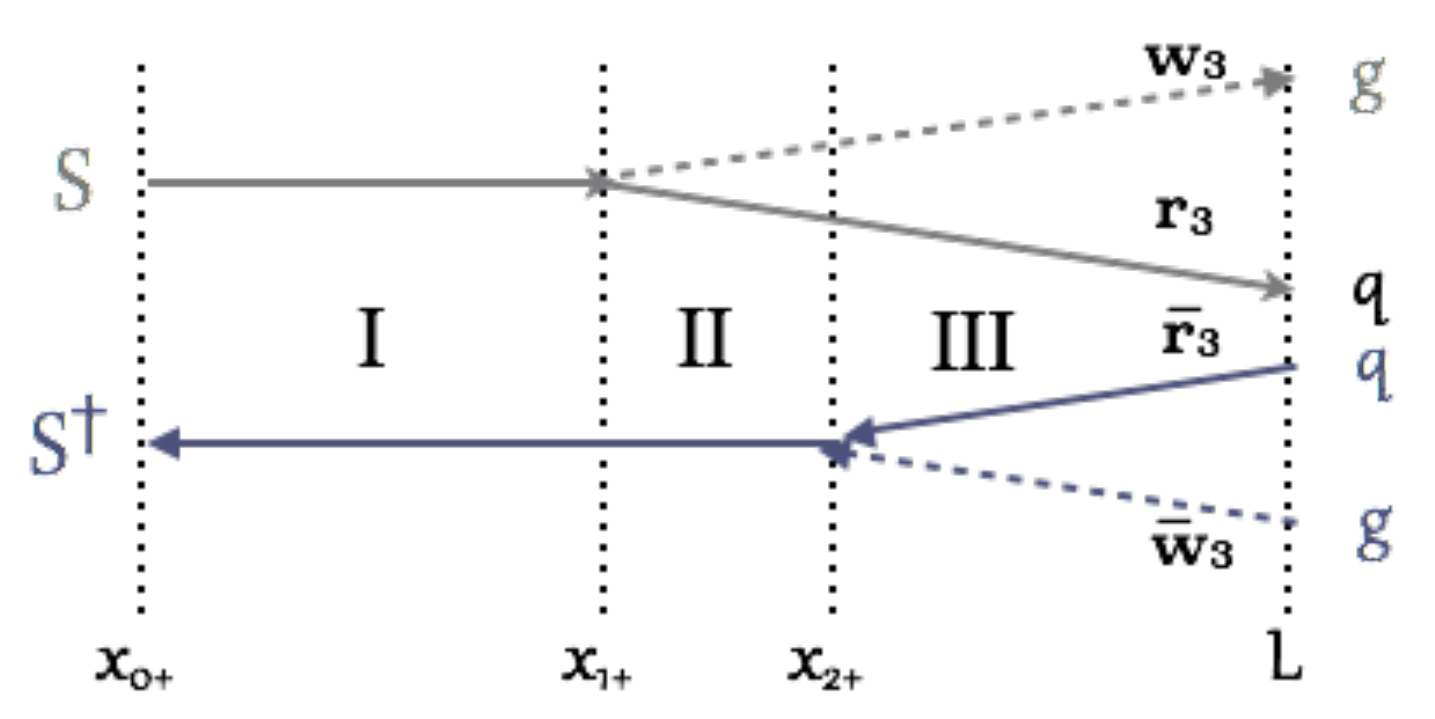}}
\caption{Schematic view of amplitude (light/grey) and conjugate amplitude (dark/purple) for the in-medium bremsstrahlung process in the presence of a finite medium. The solid lines represent a quark while the dashed lines a gluon. The transverse and (+) components of the coordinates are made explicit.}
\label{Fig:spec}
\end{figure}

The role of such results seem to be essential for a correct description of the current experimental observations. In particular, its interplay, as they were derived within different kinematic setups. Recent works in \cite{Apolinario:2014csa} have shown that, for a finite medium, within a multiple soft scattering approximation and at large $N_c$, the double differential in-medium gluon radiation spectrum, schematically represented in figure \ref{Fig:spec}, can be seen as a contribution of three factorized pieces: (i) region I, that translates the independent broadening of the initial particle, also present in \cite{Blaizot:2012fh}; (ii) region II, that exists during the formation time of the emitted gluon (and consequently neglected in \cite{Blaizot:2012fh}), whose colour correlation function in the fundamental representation translates a coherent propagation of the two emitted particles;
(iii) region III, whose colour correlation function\footnote{For results after doing all path-integrations, see \cite{Apolinario:2014csa}.} can be shown to be:
\begin{equation}
\begin{split}
	& \left\langle \text{Tr} \left[ W^\dagger (\mathbf{\bar{w}_3}) W (\mathbf{w_3}) \right] \text{Tr} \left[W^\dagger (\mathbf{w_3}) W (\mathbf{\bar{w}_3}) W^\dagger (\mathbf{\bar{r}_3}) W(\mathbf{r_3}) \right] \right\rangle \\
	& \underset{N_c \rightarrow \infty}{\simeq} \text{Tr} \left\langle W^\dagger (\mathbf{\bar{w}_3}) W (\mathbf{w_3}) \right\rangle \text{Tr} \left\langle W^\dagger (\mathbf{w_3}) W (\mathbf{\bar{w}_3}) W^\dagger (\mathbf{\bar{r}_3}) W(\mathbf{r_3}) \right\rangle \, .
\end{split}
\end{equation}
The 4-point function from above is reduced to the factorization of the independent broadening of the two final particles (as in \cite{Blaizot:2012fh}) but corrected by a decoherence parameter, $\Delta_{coh}$,
\begin{equation}
	\text{Tr} \left\langle W^\dagger (\mathbf{w_3}) W (\mathbf{\bar{w}_3}) \right\rangle \text{Tr} \left\langle W^\dagger (\mathbf{\bar{r}_3}) W(\mathbf{r_3}) \right\rangle \Delta_{coh} \,
\end{equation}
that contains an additional contribution (exponentially suppressed) in which the two particles, even after formation time, propagate coherently as in vacuum:
\begin{equation}
\begin{split}
	\Delta_{coh} = & 1 + N_c \frac{\hat{q}}{2}\int_{x_{2+}}^{L_+} d\tau \left(\mathbf{r_3}(\tau) - \mathbf{\bar{r}_3}(\tau) \right) \cdot \left(\mathbf{w_3}(\tau) - \mathbf{\bar{w}_3}(\tau) \right) \\
	& \times \exp \left\{- \frac{\hat{q}}{2} \int_{x_{2+}}^{\tau} d\xi (\mathbf{r_3} - \mathbf{w_3} )\cdot (\mathbf{\bar{w}_3} - \mathbf{\bar{r}_3}) \right\}
\end{split}
\end{equation}

As $\Delta_{coh}$ contains the same physics as $\Delta_{med}$, these results show, for the first time, the clear interplay and competition between the two propagation regimes, coherent and incoherent, when transverse Brownian motion is taken into account.

\section{Conclusions}

The jet quenching phenomenon is experimentally and theoretically established by now. While energy loss effects are in fair agreement when comparing results from several jet quenching Monte Carlos with experimental results, the description of intra-jet modifications needs to be urgently improved. Much theoretical progress, within perturbation theory, has been made in this direction with the inclusion of broadening effects and decoherence phenomena into the models. In this work, we showed a clear interplay between the two different propagation regimes (coherent and decoherent) and its relation with the medium parameters and amount of transverse momentum broadening that is acquired independently by the final particles. Therefore, it constitutes a unified and improved description of the findings derived previously in several works \cite{Apolinario:2012vy,Blaizot:2012fh,MehtarTani:2010ma}.

\textbf{Acknowldegments:} This work was supported by Fundação para a Ciência e a Tecnologia of Portugal under project CERN/FP/123596/2011. 

\bibliographystyle{plain}
\bibliography{Bibliography}


\end{document}